\documentclass[aps,prl,twocolumn,superscriptaddress]{revtex4-2} 

\usepackage{amsmath,amssymb,amsfonts} 
\usepackage{graphicx} 
\usepackage{hyperref}
\usepackage{bm} 
\usepackage{times} 
\usepackage{xcolor}
\usepackage{subcaption}
\usepackage{caption}
\DeclareCaptionStyle{mystyle}
  {format=plain,%
    textformat=period,
    justification=RaggedRight,
    singlelinecheck=true,
  }

\DeclareCaptionStyle{singlelinecentered}
  [justification=Centering]
  {style=mystyle}

\DeclareCaptionStyle{singlelineraggedleft}
  [justification=RaggedLeft]
  {style=mystyle}

\begin{document}

\title{Breakdown of adiabaticity in topological quantum liquids}

\author{Carola Ciaramelletti}
\affiliation{Dipartimento di Ingegneria e Scienze dell’Informazione e Matematica, Università dell’Aquila, via Vetoio, I-67010 Coppito-L’Aquila, Italy}
\author{Daniel Arrufat-Vicente}
\affiliation{Institut für Theoretische Physik, ETH Zürich, Wolfgang-Pauli-Strasse 27 Zürich, Switzerland}
\author{Simone Paganelli}
\affiliation{Dipartimento di Scienze Fisiche e Chimiche, Università dell’Aquila, via Vetoio, I-67010 Coppito-L’Aquila, Italy}
\author{Nicolò Defenu}
\affiliation{Institut für Theoretische Physik, ETH Zürich, Wolfgang-Pauli-Strasse 27 Zürich, Switzerland}
\affiliation{CNR-INO, Area Science Park, Basovizza, I-34149 Trieste, Italy}

\begin{abstract}
We study the temporal behavior of topological quantum fluids with strong long-range couplings under slow external perturbations, whose rate $\delta$ approaches the quasi-static limit $\delta\to 0$. As expected, due to strong long-range interactions, the system lies in the mean-field universality and the density of defects for drives across the quantum critical point is adiabatic $n_{\rm exc}\propto \delta^{2}$.  However, if the drive is instead terminated precisely at the edge of the topological non-trivial phase, the number of generated excitations becomes extensive $n_{\rm exc}\propto O(1)$. This result fundamentally breaks the established universal behavior observed in local topological quantum fluids and demonstrates a novel mechanism for the breakdown of adiabaticity in fermionic systems with strong long-range interactions.
\end{abstract}

\maketitle

\paragraph{Introduction}
Understanding the dynamical behavior of quantum many-body systems remains a fundamental challenge in modern physics\,\cite{polkovnikov2011colloquium}. In particular, systems characterized by two-body interactions that decay as a power law with distance—commonly referred to as long-range systems\,\cite{campa2014physics}—have recently obtained significant attention due to their relevance in condensed matter physics, quantum simulation, and quantum information processing\,\cite{defenu2023long}. This growing interest is largely driven by the potential to harness long-range couplings in practical quantum technologies\,\cite{guo2022implementing, solfanelli2023quantum, solfanelli2025universal,hermes2020dimensionality}. Within this context, quantum spin Hamiltonians have emerged as paradigmatic frameworks, serving as foundational models across a variety of quantum platforms\,\cite{defenu2023long}, including quantum circuits whose computational processes are often cast into effective Ising-like Hamiltonians\,\cite{kairys2020simulating, pagano2020variational, zhukov2025grover,cervera2018exact}.

Long-range interactions are closely associated with the emergence of ergodicity breaking in quantum many-body systems. Such breaking manifests in phenomena like time crystals~\cite{zhang2017observation,sacha2017time}, quantum many-body scars~\cite{turner2018weak,bluvstein2021controlling,comparin2022robust,chandran2023quantum,lerose2025theory}, weak ergodicity breaking~\cite{sorba2025quantum}, and long-lived quasi-stationary states~\cite{kastner2011diverging, defenu2021metastability, wu2023signatures}. The long-term stability of these effects is inextricably linked to strong long-range character of the system~\cite{defenu2024out}. In this context, the term \emph{strong long-range systems} typically refers to ferromagnetic quantum spin Hamiltonians with couplings that decay as a power law with distance, i.e., interactions of the form $\propto r^{-\alpha}$, where $0 \le \alpha < d$ in $d$ spatial dimensions. 

While (weak) ergodicity breaking typically follows a sudden quench, the failure of a quantum many-body system to equilibrate during a quasi-static transformation is generally referred to as adiabaticity breaking~\cite{zwerger2008limited,polkovnikov2008breakdown}. Strong long-range spin Hamiltonians are known to exhibit adiabaticity breaking~\cite{acevedo2015new}, a phenomenon associated with the emergence of a discrete quasi-particle spectrum~\cite{defenu2024out}. This spectral discreteness stabilizes collective excitations and ultimately drives the macroscopic condensation of topological defects~\cite{defenu2018dynamical,gherardini2024universal}.

The connection between adiabaticity breakdown and discrete quasi-particle spectrum is easily illustrated in the $\alpha = 0$ case. In this limit, the system reduces to an effective single-mode description, which is adiabatically driven across a critical point where its effective mass vanishes. Due to the bosonic nature of the mode, this critical point exhibits infinite degeneracy and violates the assumptions underlying the quantum adiabatic theorem~\cite{bachmann2017dynamical}. As a result, a finite density of defects persists even under perfectly quasi-static transformations~\cite{defenu2021quantum}. Thus, in quantum spin systems, a sufficiently slow decay exponent $\alpha$ stabilizes a low-energy description based on bosonic spin waves. This, in turn, when combined with a discrete quasi-particle spectrum, leads to the breakdown of adiabaticity. 

Strong long-range interactions can also coexist with fermionic excitations, as in topological quantum fluids. Indeed, it has been recently demonstrated that, despite the intrinsic connection between topological phases and locality, Majorana edge modes remain robust in the presence of strong long-range interactions~\cite{patrick2017topological}. This result opens up several questions about how long-range interactions interplay with the fermionic excitations characteristic of topological matter, setting it apart from the more traditional focus on quantum spin systems of the long-range literature.

In this context, it is important to investigate whether the hallmarks of quantum long-range physics, such as ensemble in-equivalence~\cite{defenu2024ensemble}, metastability~\cite{kastner2011diverging}, quantum scars~\cite{comparin2022robust,lerose2025theory} and adiabaticity breaking~\cite{defenu2018dynamical} continue to exist in the fermionic setting. Although pioneering studies suggest that metastability may survive in fermionic systems~\cite{defenu2021metastability,arrufatvicente2024freezing}, the breaking of adibaaticity remains more subtle as its core argument relies on the bosonic nature of the excitations in quantum magnets.

In this Letter, we intend to demonstrate that strong long-range interactions induce adiabaticity breaking also in topological quantum matter, but with a more subtle mechanism. Following Ref.~\cite{patrick2017topological} we will consider a Kitaev’s chain~\cite{kitaev2001unpaired} decorated with infinite-range tunneling and pairing couplings. At variance with Ref.~\cite{patrick2017topological}, which discussed the robustness of Majorana edge states in the open chain, we will here focus on the thermodynamic properties of the model and employ periodic boundary conditions. In order to avoid the cancellations of the infinite-range terms, we cutoff long-range interactions to half the chain length. This procedure makes the finite periodic chain easy to calculate analytically, while preserving the proper thermodynamic limit for long-range interactions~\cite{defenu2023long}.

The analysis of this model reveals solid evidence for two seemingly contradictory phenomena:
(i) In critical quenches, i.e. where the system is driven across a phase transition, strong long-range interactions tend to favor adiabatic behavior, aligning with expectations from mean-field theories and, more broadly, models above the upper critical dimension~\cite{polkovnikov2008breakdown};
(ii) At the same time, these strong long-range interactions give rise to a novel non-adiabatic dynamical phase, fundamentally distinct from its bosonic counterpart~\cite{defenu2018dynamical}.

Our findings confirm that strong long-range interactions inherently drive non-adiabatic dynamics. Remarkably, in topological fluids, this behavior can be demonstrated beyond the singular case of $\alpha = 0$ to all $\alpha < d$, reinforcing the expectations of universality also for generic quantum magnets~\cite{defenu2021quantum}.

\paragraph{Decorated Kitaev chain:}
As anticipated, we consider the out-of-equilibrium dynamics of an analytically solvable model of a topological fluid initially at zero temperature, when the chemical potential is linearly ramped across the equilibrium phase boundaries. The study examines the Kitaev chain decorated with long-range interactions:
\begin{align}
    \hat{H}= -\sum_{i=1}^{N} \big{[}\sum_{r=1}^{N/2-1} (j_r \hat{c}^{\dag}_i \hat{c}_{i+r} + \Delta_r \hat{c}^{\dag}_i \hat{c}^{\dag}_{i+r} + h.c) \nonumber
    \\
    -2 \mu \hat{c}^{\dag}_i \hat{c}_i \big{]}\,,
    \label{Kit}
\end{align}
where $j^{\alpha}_r = \frac{\lambda}{N_{\alpha}} \frac{1}{r^{\alpha}}$ is the hopping term, $\Delta^{\beta}_r = \frac{\lambda}{N_{\beta}} \frac{1}{r^{\beta}}$ is the pairing term $N$ is the number of sites and $\mu$ is the on-site potential. The coupling strengths of the pairing and hopping terms do not play any role in our study and they have been both fixed to a single value $\lambda$. We will focus on the strong long-range regime $\alpha=\beta <1$, where Kac scaling $N_{\gamma}^{-1} \sim (1- \gamma) 2^{(1-\gamma)} N^{\gamma -1}$ is employed~\cite{defenu2024out}.

 The translational invariance of the Hamiltonian makes it convenient to express the system in Fourier space
\begin{align}
\hat{b}_k = \frac{e^{-i\pi/4}}{\sqrt{N}} \sum_{r \in \mathbf{Z}} \hat{c}_r e^{-ikr},
\end{align}
where the momentum values are discretized according to the boundary conditions: for periodic boundary conditions (PBC), they are given by \( k = \frac{2\pi n}{N} \), while for antiperiodic boundary conditions (ABC), they take the form \( k = \frac{2\pi n + \pi}{N} \), with \( n \in \mathbf{Z} \) and \( -\frac{N}{2} \leq n < \frac{N}{2} \). Given the strong long-range nature of the couplings the spectrum of the matrices $j_{r}$ and $\Delta_{r}$ will remain discrete also in the thermodynamic limit~\cite{defenu2021metastability}. We will use the integer index $n$ as a label for Fourier modes from now on.
\begin{figure}[t]
    \includegraphics[width=0.5\textwidth]{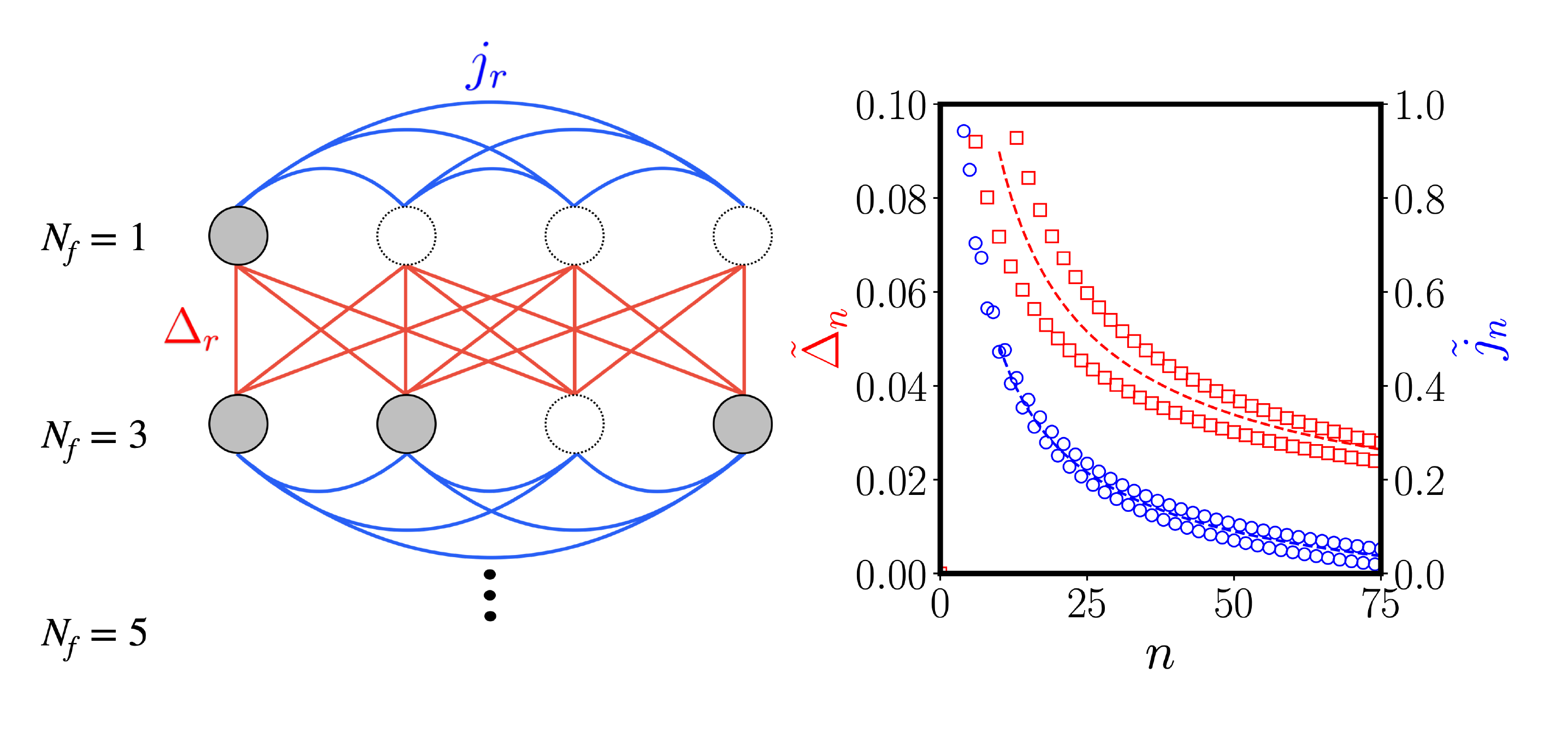}
  \centering
    \caption{The hopping (blue) and pairing (red) couplings of the strong long-range Kitaev chain. Panel a) illustrates both the single particle hopping terms ($j_{r}\hat{a}^{\dagger}_{i+r}\hat{a}_{i}$), represented by the blue lines, as well as the pairing terms ($ \Delta \hat{a}^{\dagger}_{i+r}\hat{a}^{\dagger}_{i}  + \text{h.c.}$) depicted by the red ones. The hopping term allows particles (grey shaded dots) to hop in between sites while the pairings allows for processes where the particles are created or annihilated in pairs, increasing or decreasing the total number of fermions $N_f$ by two ($N_f\to N_f \pm2$). Panel b) represents the result of the Fourier series for a finite system of size $N=2^{10}$, $\alpha=2/\pi$ and $\lambda=1$. The dashed lines are the asymptotic expressions obtained in the thermodynamic limit (see Eq.\,\eqref{kin_en} and Eq.\,\eqref{pair_en})}
    \label{Fig1}
\end{figure}

The Hamiltonian can, then, be expressed in terms of momentum modes as:
\begin{equation}
\label{momentum_space_hamiltionian}
\hat{H} = \sum_n \Big[2\hat{b}_n^{\dag} \hat{b}_n \varepsilon_{n}(t) + (\hat{b}_{n}^{\dag} \hat{b}_{-n}^{\dag} + h.c.) \tilde{\Delta}_{n} \Big],
\end{equation}
with $\varepsilon_{n}(t) = \mu(t) - \tilde{j}_{n}$. The momentum space couplings read
\begin{align}
&\tilde{j}_{n} = \frac{\lambda }{N_{\alpha}} \sum_{r=1}^{\frac{N}{2}-1} \frac{\cos(kr)}{\left(\frac{r}{N}\right)^{\alpha}}\stackrel{n\gg1}{\approx}\frac{s_\alpha \lambda}{n^{1-\alpha}},\label{kin_en}\\
&\tilde{\Delta}_{n} = \frac{\lambda }{N_{\alpha}} \sum_{r=1}^{\frac{N}{2}-1} \frac{\sin(kr)}{\left(\frac{r}{N}\right)^{\alpha}}\stackrel{n\gg1}{\approx} \frac{c_\alpha \lambda}{n^{1-\alpha}}\label{pair_en}.
\end{align}
where $s_\alpha=\sin (\alpha \pi / 2) \Gamma(2-\alpha)(2 \pi)^{\alpha-1}$ and $c_\alpha=\cos (\alpha \pi / 2) \Gamma(2-\alpha)(2 \pi)^{\alpha-1}$. The expressions on the r.h.s. of the $\approx$ symbol represent asymptotic expansions at large-$n$ obtained discarding respectively terms of order $\mathcal{O}\left(n^{ \alpha-3/2}\right)$ and $\mathcal{O}\left(n^{-1}\right)$. The vanishing of the spectral couplings at large-$n$ is a consequence of strong long-range interactions, see Fig.~\ref{Fig1}.

The Hamiltonian in Eq.~\eqref{momentum_space_hamiltionian} 
can be diagonalized via the Bogoliubov transformation 
$\hat{b}_k=u_k(t) \hat{\gamma}_k+v_{-k}^*(t) \hat{\gamma}_{-k}^{\dagger}$~\cite{chakrabarti1996transverse}. 
The resulting quasi-particle spectrum becomes gapless at the critical point 
$\mu_c = \lambda$, independent of the values of $\alpha$ and $\beta$. 
This marks the onset of a quantum phase transition, separating a topologically 
trivial phase for $\mu > \lambda$ from a topologically non-trivial phase for 
$0<\mu < \lambda$~\cite{solfanelli2023logarithmic} 
\footnote{The topologically non-trivial phase regime depends on $\alpha$ in the following manner: 
For $0<\alpha<1$ the regime corresponds to $0<\mu<\lambda$. For $1<\alpha<2$ to $\lambda\left(-1+2^{1-\alpha}\right)<\mu<\lambda$ and for $\alpha>2$ to $-\lambda<\mu<\lambda$}. 
The emergence of a non-trivial topological index 
is associated with the presence of edge modes 
in systems with open boundary conditions~\cite{patrick2017topological,solfanelli2023logarithmic}.

\emph{Quasi-static dynamics:} We consider a global quasi-static transformation in which the chemical potential varies slowly across the quantum critical point at $\mu_{c}=\lambda$. Due to the universal behavior of systems near phase transition, one can, without loss of generality, consider a linear ramp of the chemical potential given by  
$\mu(t) = \lambda-\delta t,$  
where $t \in \left[t_{\mathrm{in}}, t_f\right]$. More generic ramps with $\mu(t)=f(\delta t)$ can be reduced to the previous case in the quasi-static limit $\delta\to 0$. It is convenient to parametrize the extrema of the ramp as $t_{\mathrm{in}}=-\lambda/\delta$ and $t_{f}=\lambda_f/\delta$. The system is initially prepared in its ground state at $t_{\mathrm{in}}$ and evolves according to the time-dependent Schrödinger equation. When reformulated in terms of the time-dependent Bogoliubov amplitudes $u_n(t)$ and $ v_n(t)$, the system dynamics takes the well-known Landau-Zener form~\cite{dziarmaga2005dynamics}
\begin{align}
    i \frac{\mathrm{~d}}{\mathrm{~d} t}\binom{u_n}{v_n}=\left[\begin{array}{cc}\varepsilon_{n}(t) & \tilde{\Delta}_{n} \\  \tilde{\Delta}_{n} & -\varepsilon_{n}(t)\end{array}\right]\binom{u_n}{v_n}.
    \label{lz_model}
\end{align}

\emph{Dynamical phase diagram:} Dynamical phases are characterized by the defect distribution $p_n(t)=|\langle _{n}\text{E.S.}| \psi(t)\rangle|^{2}$, where $_{n}\langle \text{E.S.}|$ corresponds to the
instantaneous excited eigenstate of the $n$ momentum mode, and $|\psi(t)\rangle$ is the time-evolved initial zero temperature state after being quenched. In this context, it is convenient to define an effective temperature by assuming a Dirac-Fermi distribution for the quasi-particle population after the quench: $p_n(t)=[1+\mathrm{exp}(\varepsilon_{n}/T_{n})]^{-1}$, where $\varepsilon_{n}$ is the kinetic energy at the beginning of the time evolution and $T_{n}$ denotes the effective post-quench temperature. Depending on the final coupling $\lambda_{f}$, the temperature profile reaches its maximum at a different quasi-particle index $n$. For quenches within the trivial phase, where $\lambda_{f}>\lambda$, the maximum temperature occurs away from the low-energy region, leaving the critical mode unexcited, $T_{n=0}\to 0^{+}$. This regime is referred to as the ``\emph{zero-temperature phase}''.
\begin{figure*}[t]
    \centering
    \begin{subfigure}[b]{0.33\textwidth}
    \centering
        \includegraphics[width=1.\textwidth]{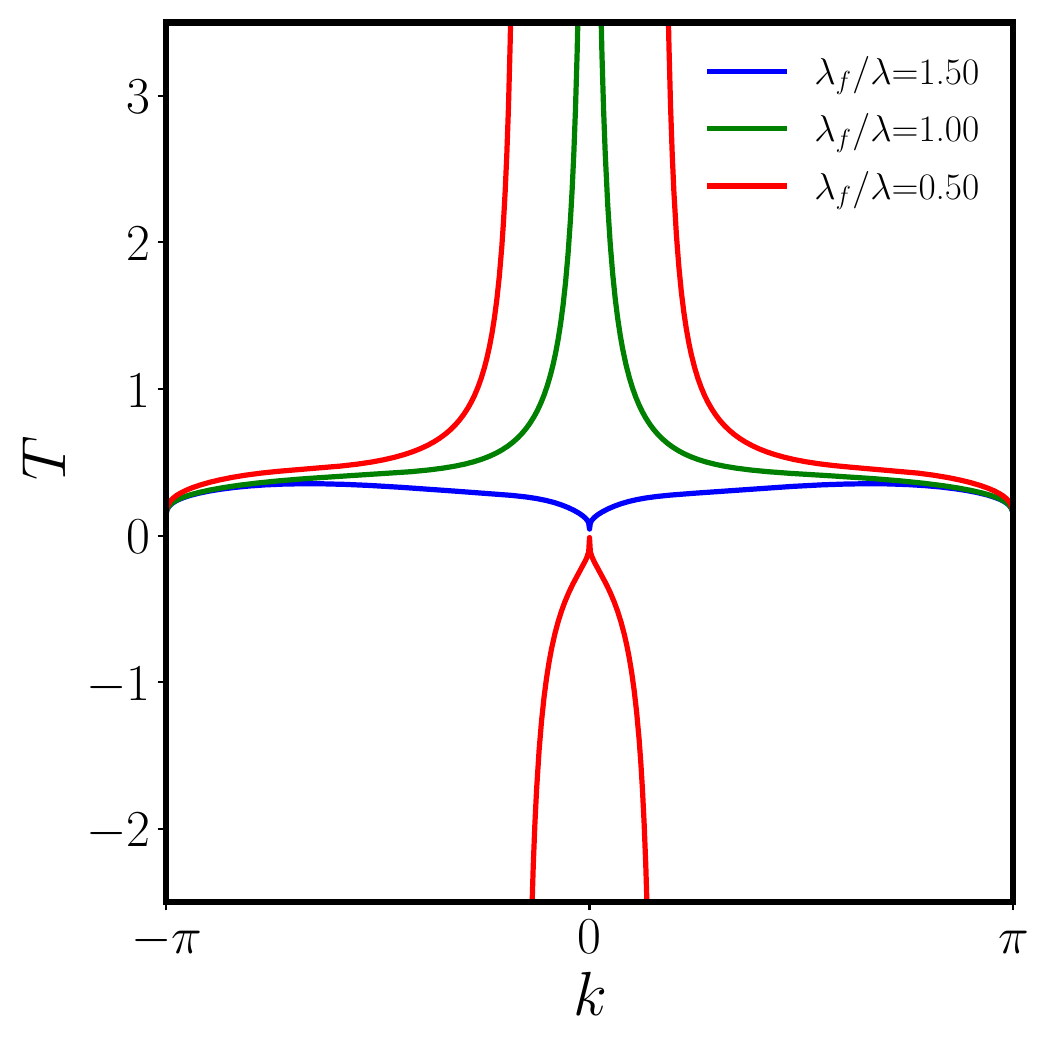}
    \end{subfigure}
    \begin{subfigure}[b]{0.33\textwidth}
    \centering
        \includegraphics[width=1.\textwidth]{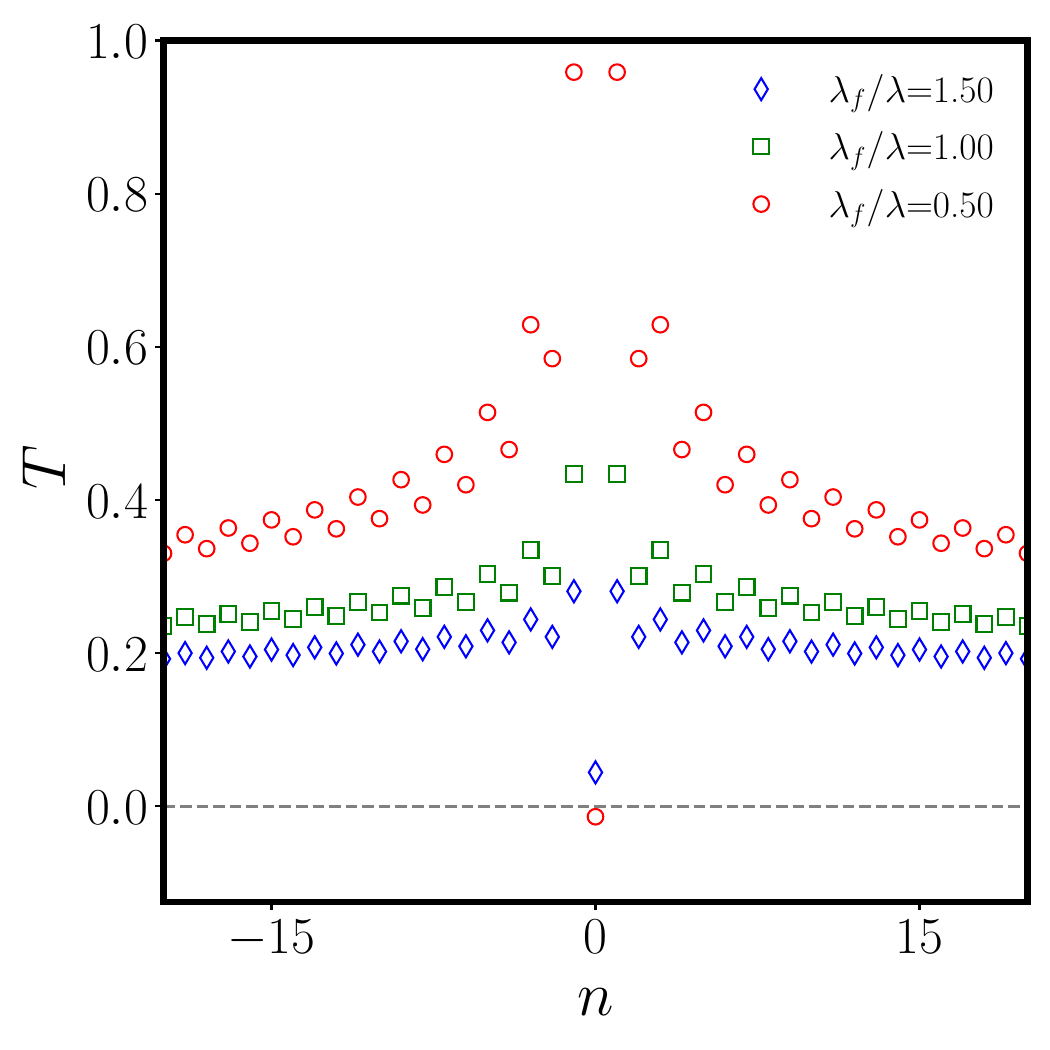}
    \end{subfigure}
       \begin{subfigure}[b]{0.33\textwidth}
    \centering
        \includegraphics[width=1.\textwidth]{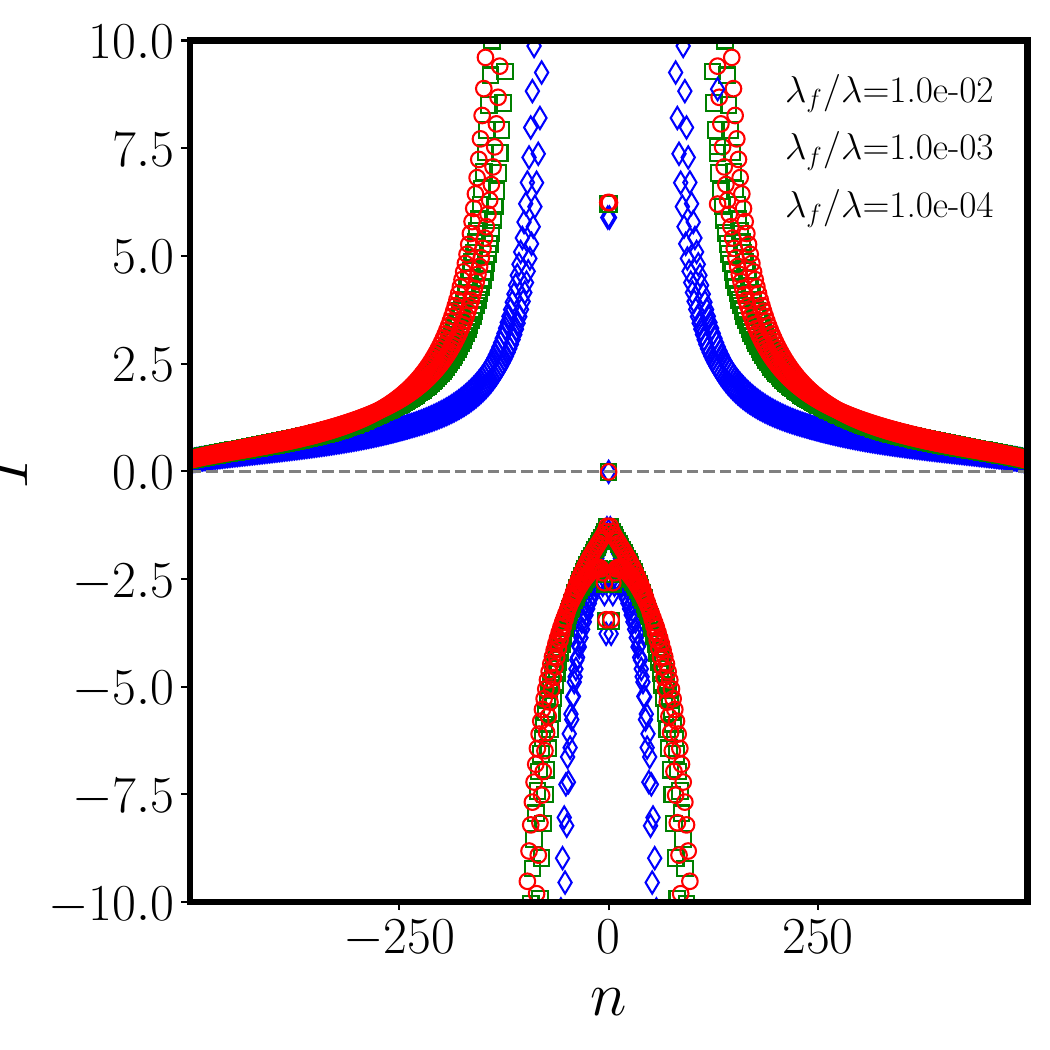}
    \end{subfigure}
    \caption{\label{Fig2}The temperature profiles in the three out-of-equilibrium phases—positive temperature (blue), negative temperature (red), and infinite temperature (green)—are shown as functions of momentum $k$ in the short-range case ($\alpha \gtrsim 11$, left panel) and as functions of mode index $n$ in the strong long-range case ($\alpha \simeq 0.5$, center panel). The fluctuations observed in the strong long-range case vanish in the thermodynamic limit. The right panel shows the increase of the negative  temperature region as the endpoint of the quench goes deeper in the broken phase $\lambda_{f}\to 0$.  All lines have been computed for $N=1024$ and $\delta=1$}
\end{figure*}

The \emph{``infinite temperature phase''} emerges as the final coupling approaches the critical value, $\lambda_f \to \lambda$. In this regime, the defect distribution becomes sharply peaked at the critical mode $n = 0$, with $p_{n=0} = 1/2$, causing the effective temperature profile to diverge. This behavior is illustrated in Fig.~\ref{Fig2}. When the quasi-static transformation crosses the critical coupling, i.e. $\lambda_f < \lambda$, the temperature of the low-energy modes becomes negative, indicating a population inversion of quasi-particles near the critical mode, where $p_{n=0} = 1$. We refer to this regime as the \emph{``negative temperature phase''}.

The three phases are distinguished by the behavior of the temperature profile near the lowest eigenmode, $n=0$. Owing to universality, the existence and structure of these temperature singularities are independent of the specific values of the final coupling $\lambda_f$ and the ramp speed $\delta$. In fact, all three phases persist for any $\alpha$, including the local limit $\alpha \to \infty$. However, for $\alpha > d$, the temperature curve remains continuous, as shown in Fig.~\ref{Fig2}, left panel, whereas in the strong long-range regime the spectrum consists only of discrete eigenmodes, as illustrated in Fig.~\ref{Fig2}, central panel. This fundamental difference leads to markedly distinct physical behavior in the two cases.

In the quasi-static limit ($\delta\to 0$) the dominant contribution to the defect distribution for critical quenches with $\lambda_{f}\geq \lambda$ takes the typical Landau-Zener form\,\cite{dziarmaga2010dynamics}
\begin{equation}
\label{LZ_formula}
    p_n =\exp \left( -\frac{\pi \tilde{\Delta}_{n}^2}{\delta} \right)+O(\delta^{2}),
\end{equation}
In the local case ($\alpha>2$), the nonanalytic scaling of the defect density predicted by the Kibble-Zurek mechanism (KZM)~\cite{zurek2005dynamics} can be recovered by computing the average occupation probability, $n_{\mathrm{exc}} = \sum p_n / N$. As $N$ increases, the sum over discrete energy levels approaches a continuous integral, and the characteristic nonanalytic KZM scaling, $n_{\mathrm{exc}} \propto \sqrt{\delta}$, emerges from integrating across the $k \approx 0$ singularity~\cite{dziarmaga2005dynamics}. A similar behavior is observed throughout the entire weak long-range regime, $\alpha, \beta > 1$, where modifications to the scaling at the singularity may also affect the scaling of the defect density~\cite{defenu2019universal}.

In the strong long-range regime, $\alpha, \beta < 1$, the excitation probability $p_n$ never approaches a continuous function. As a result, the low-energy region at $n \ll N$, despite exhibiting the highest excitation probabilities, does not become dense. In other words, this region has zero measure in the thermodynamic limit, and its contribution to the overall excitation density vanishes. At $\alpha, \beta < 1$, the only momentum region that becomes dense in the thermodynamic limit is located at large $n$, where states accumulate near a single point as $n \to \infty$.

However, the modes close to the accumulation point lie beyond the applicability of Eq.~\eqref{LZ_formula}. Indeed, the Landau-Zener formula\,\eqref{LZ_formula} applies only when the system is driven through an avoided energy crossing. This condition is met only for those quasi-particles whose kinetic energy changes sign during the dynamics. Since $\varepsilon_n(t_{\mathrm{in}}) > 0$ for all $n$, the avoided crossing condition reduces to
\begin{equation}
\label{nmax}
    \varepsilon_n(t_f) < 0 \quad \Rightarrow \quad n < n_{\mathrm{max}} \approx \left( \frac{s_{\alpha}}{\lambda_f} \right)^{\frac{1}{1 - \alpha}},
\end{equation}
where the final expression is obtained using the asymptotic expansion on the right-hand side of Eq.\,\eqref{kin_en}. Thus, only quasi-particles with $n < n_{\mathrm{max}}$ satisfy the conditions for applying Eq.\,\eqref{LZ_formula}, which corresponds to an $\mathcal{O}(1/N)$ fraction of the spectrum. In the dense region at $n \gg n_{\mathrm{max}}$, the excitation probability must be computed by solving the Landau-Zener problem far from the avoided crossing.

The Landau-Zener problem can be solved analytically for arbitrary values of $t_{\mathrm{in}}$ and $t_f$. However, to determine the scaling of the excitation probability with respect to the ramp speed $\delta$, one can employ adiabatic perturbation theory, which shows that $p_{n > n_{\mathrm{max}}} \propto \delta^2$\,\cite{degrandi2010adiabatic}. This immediately leads to the scaling $n_{\mathrm{exc}} \propto \delta^2$, even within the infinite and negative temperature phases at least as long as $\lambda_{f}>0$. The absence of the nonanalytic Kibble-Zurek scaling for the defect density in the strong long-range interaction regime is consistent with the effective dimension approach. This framework predicts that the system lies above the upper critical dimension\,\cite{solfanelli2025universal}, resulting in the adiabatic scaling $n_{\mathrm{exc}} \propto \delta^2$\,\cite{polkovnikov2005universal}.

 \paragraph{The non-adiabatic phase:} In strong long-range topoligical quantum fluids adiabaticity breaking can only occur in the thermodynamic limit if the high energy modes close to the degenerate point achieve infinite temperatures. This corresponds to having $n_{\mathrm{max}}\to\infty$, which, based on Eq.\,\eqref{nmax}, corresponds to quenches with $\lambda_{f}\to 0$. This effect is readily observed studying the temperature profiles for slow quenches at vanishing $\lambda_{f}=[10^{-2},10^{-3},10^{-4}]$, see the right panel in Fig.\,\ref{Fig2}, where it is shown that negative temperatures are achieved for an increasingly larger portion of the spectrum as $\lambda_{f}$ decreases.

To understand the physics of the non-adiabatic phase at $\lambda_{f}=0$ it is convenient to rewrite Eq.\,\eqref{lz_model} in the asymptotic form
\begin{align}
\label{largeNeffectivemodel}
    i \frac{\mathrm{~d}}{\mathrm{~d} t}\binom{u_n}{v_n}=\left[\begin{array}{cc}\mu'(t)-\frac{s_\alpha \lambda}{n^{1-\alpha}} & \frac{c_\beta \lambda}{n^{1-\alpha}} \\  \frac{c_\beta \lambda}{n^{1-\alpha}} & -\mu'(t)+\frac{s_\alpha \lambda}{n^{1-\alpha}}\end{array}\right]\binom{u_n}{v_n},
\end{align}
which has been obtained by substituting the asymptotic form of the momentum space couplings on the r.h.s. of Eqs.\,\eqref{kin_en} and\,\eqref{pair_en} into the e.o.m. in Eq.\,\eqref{lz_model}. 
The drive $\mu'(t)=-\delta t$ with $t\in [-t/\delta,0]$ has been chosen so that the critical point of the high-energy mode is approached as $t\to 0^{-}$ \footnote{Equivalently, one can take the limit $t\in [-\infty,0]$ when studying the $\delta \to 0$ limit}. Indeed, since $\tilde{\Delta}_{n\to\infty}\to0$ large-$n$ excitations effectively become critical at $\mu=0$ and the slow quench with $\lambda_{f}=0$ corresponds to a slow ramp to the critical point. 

\begin{figure}[h]
    \includegraphics[trim={0.cm 0.cm 0.cm 0.cm},clip,width=0.45\textwidth]{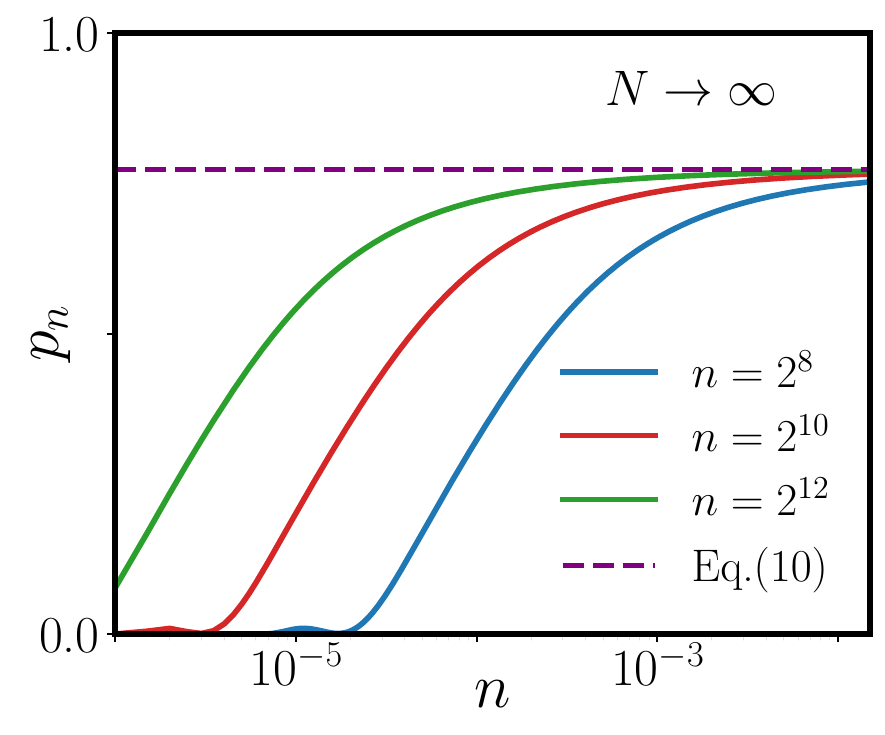}
  \centering
    \caption{Behavior of $p_{n}$ as a function of $\delta$ for the effective large-$N$ model in Eq.~\eqref{largeNeffectivemodel} with $t_{\text{f}}=0$. The behavior of different $n$ is shown and in all cases $\alpha=1/e$ as well as the theoretical prediction from Eq.\eqref{eq:thermodynamiclimitfirst}.  We can observe how, as we approach the high energy modes ($n\approx N$) the excitation probability exactly corresponds to the constant value predicted by Eq.\eqref{eq:thermodynamiclimitfirst}}
    \label{Fig3}
\end{figure}

When approaching this critical point not only the gap approaches zero, $\tilde{\Delta}_{n\to\infty}\to0$, but also the fermionic dispersion relation $\varepsilon_{n\to\infty}(t_{\text{fin}})\to0$ does. 
As we show in the Supplemental material \footnote{See Supplemental material I.} this drastically changes the behavior of the half-ramp dynamics compared to the prototypical Landau-Zener solution. For the high energy modes $n$, when taking the thermodynamic limit $N\to \infty$, the excitation probability becomes independent of $n$ and converges to a constant value
\begin{equation}\label{eq:thermodynamiclimitfirst}
    p_{n}=\sin^2\left(\frac{\pi (1+\alpha)}{4}\right) + \mathcal{O}(N^{\alpha-1})
\end{equation}
for any finite $\delta$. However, when the adiabatic limit, $\delta \to 0$, is taken before the thermodynamic limit, $N \to \infty$ the adiabatic scaling $n_{\rm exc}\sim \delta^{2}$ is recovered. In Fig.\ref{Fig3} we can observe how the probability of finding the high energy modes in their corresponding excited state interpolates between the exponential behavior of the conventional full-ramp Landau-Zener $p_{\text{exc}}=\exp\left[-\pi \tilde{\Delta}^2_n/4\delta\right]$ \cite{polkovnikov2008breakdown} and the constant value reported in Eq.\,\eqref{eq:thermodynamiclimitfirst}.
 The number of excitations is defined as the sum of excitation over each mode, which, for the case $\alpha<1$ and, because of the extensive number of high energy modes, corresponds to
 \begin{equation}
    N_{\rm exc} = \sum_{k} p_{k}\propto C_{\alpha}\frac{N}{2}\left(\sin^2\left(\frac{\pi (1+\alpha)}{4}\right)\right).
\end{equation}
where $C_{\alpha}\in (0,1)$ accounts for the fact that the high energy modes only represent a finite fraction of the overall spectrum which we expect to depend on $\alpha$.
Interestingly, the number of excitations is extensive, in contradiction with the typical Kibble-Zurek scenario where the defect density vanishes as $\delta\to 0$ for thermodynamically large systems~\cite{dziarmaga2010dynamics} and in contradiction to other models that also present topological defects where the defect scaling is proportional to the area of the sample \cite{uhlmann2007vortex,uhlmann2010ONsymmetry}. A similar effect is known to happen for bosonic systems~\cite{defenu2018dynamical}. Here, at variance from the bosonic case, where the low energy mode can become macroscopically occupied, each energy mode can only be occupied by a single fermion. However, the accumulation of the high energy modes near the same point leads to a macrocscopic contribution to the defect probability.

 We test Eq.~\eqref{eq:thermodynamiclimitfirst} against the numerical results obtained by explicitly solving the dynamical evolution of the effective large-$N$ model in Eq.~\eqref{largeNeffectivemodel}.  For the ramp $\mu'(t)$ the excitation probability becomes independent of $\delta$ for large enough $n$. The asymptotic value at $\delta\gg 1/n^{2\alpha-2}$ coincides with Eq.~\eqref{eq:thermodynamiclimitfirst}, which is shown as a black dashed line in Fig.~\ref{Fig3}.

\emph{Suppressed tunneling:}  Our study shows that the adiabatic theorem fails for topological quantum liquids with $\alpha=\beta<1$ because level crossings become unavoidable. This breakdown is best understood through the Landau–Zener tunneling amplitude $\tilde{\Delta}_{n}$, which vanishes for high-energy states in the strong long-range regime. As tunneling is suppressed, level switching is governed solely by the system’s dynamical evolution.
In conventional quantum systems, tunneling enables hybridization between classically distinct levels, preserving adiabaticity within the Landau–Zener framework. 

By contrast, in the strong long-range regime, the energy gaps between high-energy modes asymptotically vanish as these modes become highly degenerate and diverge in number. With no tunneling, Landau–Zener transitions occur between classical states, and adiabaticity breaks down. Since the tunneling amplitude only vanishes at large $n$, this breakdown is not universal. It appears exclusively at $\lambda_{f}=0$, far from the low-energy critical point $\lambda_{f}=\lambda$, where defect-density scaling remains adiabatic.

\paragraph{Conclusions:} In this work, we have analyzed the nonadiabatic nature of long-range topological quantum fluids by studying the dynamics of a solvable fermionic model, the Kitaev chain with long-range hopping and pairing terms. Our results demonstrate that strong long-range interactions inherently lead to a breakdown of adiabaticity also in systems with fermionic quasi-particles, extending previous findings on quantum ferromagnets at $\alpha=0$~\cite{acevedo2015new, gherardini2024universal, defenu2018dynamical}. 

A key outcome of our study is that in the strong long-range regime, Landau-Zener tunneling is suppressed for all high-energy modes, which become increasingly degenerate and ultimately behave as classical states. This results in a level switching mechanism that is dictated purely by the classical dynamics rather than by quantum tunneling effects.

Long-range interactions frequently appear in the context of quantum thermodynamics and quantum annealing applications \cite{solfanelli2025universal, bode2024adiabatic}. Our findings highlight that strong long-range interactions may introduce fundamental limitations to adiabatic quantum state preparation. Their nature suppresses tunneling at high energy and prevents smooth transitions between eigenstates, which is a crucial requirement for adiabatic quantum computation and annealing. 

The inclusion of local quartic terms in Eq.~\eqref{momentum_space_hamiltionian} is expected to have a minimal impact and does not alter our main conclusions. Indeed, the gapped nature of quasi-particles, combined with Fermi exclusion statistics, strongly suppresses interaction effects during sudden quenches, regardless of quench strength~\cite{arrufatvicente2024freezing}. This contrasts with quantum spin Hamiltonians, where the suppression of many-body effects is limited to scar states within the spectrum~\cite{lerose2025theory}. An extension of the latter studies to the case of quasi-static dynamics is envisaged.

\emph{Acknowledgements:}
 This research was funded by the Swiss National Science Foundation (SNSF) grant numbers 200021--207537 and 200021--236722, by the Deutsche Forschungsgemeinschaft (DFG, German Research Foundation) under Germany's Excellence Strategy EXC2181/1-390900948 (the Heidelberg STRUCTURES Excellence Cluster) and by the European Union under GA No. 101077500–QLRNet. Partial support by grant NSF PHY-230935 to the Kavli Institute for
Theoretical Physics (KITP) is also acknowledged.
\label{appendix}

\onecolumngrid 
\vspace{1em}
\section{Supplemental Material}
\twocolumngrid 
\emph{The Landau Zenner problem:} Here we revise the derivation for the exact solution to the time-dependent Bolgoliubov rotation which is equivalent to solving the transition probabilities in the Landau-Zener (LZ) problem. The equation to solve is 
\begin{align}\label{eq_Bogoliubov_time}
    i \frac{\mathrm{~d}}{\mathrm{~d} t}\binom{u_n}{v_n}=
    \left[\begin{array}{cc}\varepsilon_{n}(t) & \tilde{\Delta}_{n} \\  \tilde{\Delta}_{n} & -\varepsilon_{n}(t)\end{array}\right]\binom{u_k}{v_k},
\end{align}
where $\tilde{\Delta}_n$ is the BCS gap of for the corresponding energy level $n$
and  $\varepsilon_n(t)=\mu(t)-\tilde{j}_n=\lambda-\delta t - \tilde{j}_n$ is the fermionic dispersion relation. 
By shifting the time $t \rightarrow t-t_n$, where $t_n\equiv (\lambda-\tilde{j}_n)/\delta$, and a posteriori re-scaling of $t=t'/\sqrt{2\delta}$
we can simplify the system of differential equations 
and find an exact solution in terms of $t'$
\begin{widetext}
    \begin{eqnarray}
        v_n(t')&=&c_1e^{-i{t'}^2/4}M\left(\frac{i\tilde{\Delta}^2_n}{4\delta},\frac{1}{2},\frac{i{t'}^2}{2}\right)
        -i
        c_2 \frac{\tilde{\Delta}_n}{\sqrt{2\delta}} t' e^{-i{t'}^2/4}M\left(\frac{1}{2}+\frac{i\tilde{\Delta}^2_n}{4\delta},\frac{3}{2},\frac{i{t'}^2}{2}\right),
        \\
            u_n(t')&=&c_2e^{-i{t'}^2/4}M\left(\frac{1}{2}+\frac{i\tilde{\Delta}^2_n}{4\delta},\frac{1}{2},\frac{i{t'}^2}{2}\right)
        -i
        c_1 \frac{\tilde{\Delta}_n}{\sqrt{2\delta}} t' e^{-i{t'}^2/4}M\left(1+\frac{i\tilde{\Delta}^2_n}{4\delta},\frac{3}{2},\frac{i{t'}^2}{2}\right).
    \end{eqnarray}
\end{widetext}
Here $M(a,b,z)$ is the confluent hypergeometric function
sometimes referred to as ${}_1F_1$.

The coefficients $c_1$ and $c_2$ are determined by the initial conditions. Using the asymptotic expansion
\begin{eqnarray}\label{eq:confluent_expansion}
    \lim_{|z|\rightarrow \infty} M(a,b,z) \approx \Gamma(b)\left(\frac{e^z z^{a-b}}{\Gamma(a)}+\frac{(-z)^{-a}}{\Gamma(b-a)}\right),
\end{eqnarray}
we can fix the coefficients $c_1$ and $c_2$ by considering the case where we start in the ground state at $t_{\text{in}}=-\infty$. This corresponds to fixing $v_n(-\infty)=1$ and $u_n(-\infty)=0$
which itself determines the values of $c_1$ and $c_2$
up to an irrelevant overall phase factor
\begin{widetext}
    \begin{eqnarray}
        v_n(t')&=&\frac{ \sqrt{\pi} e^{-\frac{\pi \tilde{\Delta}^2_n}{8\delta}}}{\Gamma\left(\frac{1}{2}+\frac{i\tilde{\Delta}^2_n}{4\delta}\right)}
        e^{-i{t'}^2/4}M\left(\frac{i\tilde{\Delta}^2_n}{4\delta},\frac{1}{2},\frac{i{t'}^2}{2}\right)
        + e^{-\frac{i5\pi}{4}}\frac{\tilde{\Delta}^2_n}{2\delta}
        \frac{ \sqrt{\pi} e^{-\frac{\pi \tilde{\Delta}^2_n}{8\delta}}}{\sqrt{2} \Gamma\left(1+\frac{i\tilde{\Delta}^2_n}{4\delta}\right)} 
        t' e^{-i{t'}^2/4}M\left(\frac{1}{2}+\frac{i\tilde{\Delta}^2_n}{4\delta},\frac{3}{2},\frac{i{t'}^2}{2}\right) ,\label{eq:vn_zener}
        \\
        u_n(t')&=&e^{-\frac{i3\pi}{4}}\frac{\tilde{\Delta}_n}{\sqrt{2\delta}}
        \frac{ \sqrt{\pi} e^{-\frac{\pi \tilde{\Delta}^2_n}{8\delta}}}{\sqrt{2} \Gamma\left(1+\frac{i\tilde{\Delta}^2_n}{4\delta}\right)}  
        e^{-i{t'}^2/4}M\left(\frac{1}{2}+\frac{i\tilde{\Delta}^2_n}{4\delta},\frac{1}{2},\frac{i{t'}^2}{2}\right)\label{eq:un_zener}
        \\
        &&\nonumber \hspace{6cm} + e^{\frac{-i\pi}{2}}
        \frac{ \sqrt{\pi} e^{-\frac{\pi \tilde{\Delta}^2_n}{8\delta}}}{\Gamma\left(\frac{1}{2}+\frac{i\tilde{\Delta}^2_n}{4\delta}\right)} \frac{\tilde{\Delta}_n}{\sqrt{2\delta}} t' e^{-i{t'}^2/4}M\left(1+\frac{i\tilde{\Delta}^2_n}{4\delta},\frac{3}{2},\frac{i{t'}^2}{2}\right) .
\end{eqnarray}
\end{widetext}

To compute the excitation probability, we need to compute the overlap between our time-evolved state $|\psi(t)\rangle$ and the instantaneous excited state $|E.S\rangle_n $
\begin{eqnarray}
    |\text{E.S.}(t) \rangle_n=\cos\left[\frac{\theta_n(t)}{2}\right]| 1\rangle + \sin\left[\frac{\theta_n(t)}{2}\right]| 2\rangle \,,
\end{eqnarray}
where the states $|0\rangle$ and $|1\rangle$ represent the empty or occupied momentum mode $n$ respectively and $\theta_n(t)$ corresponds to the Bogoliubov rotation at time $t$ 
\begin{eqnarray}
    \cos\left[\frac{\theta_n(t)}{2}\right]&=&\left(\frac{\left(\varepsilon^2_n(t)+\tilde{\Delta}^2_n\right)^{1/2}-\varepsilon_n}{2\left(\varepsilon^2_n(t)+\tilde{\Delta}^2_n\right)^{1/2}}\right)^{1/2} ,
\end{eqnarray}
and
\begin{eqnarray}
    \sin\left[\frac{\theta_n(t)}{2}\right]&=&\left(\frac{\left(\varepsilon^2_n(t)+\tilde{\Delta}^2_n\right)^{1/2}+\varepsilon_n}{2\left(\varepsilon^2_n(t)+\tilde{\Delta}^2_n\right)^{1/2}}\right)^{1/2}.
\end{eqnarray}
The excitation probability is defined as the overlap between the exact time-evolved state and the instantaneous excited state $p_n(t)\equiv |{}_n\langle \text{E.S.}| \psi(t)\rangle|^{2}$ is computed using
\begin{eqnarray}
    p_n(t)=\left|\cos\left(\frac{\theta_n(t)}{2}\right)v_n(t)+\sin\left(\frac{\theta_n(t)}{2}\right)u_n(t)\right|^2 \,.\,\,\,\,
\end{eqnarray}
Note how the corresponding $t'$ from Eq.\eqref{eq:vn_zener} and Eq.\eqref{eq:un_zener} is set to $t'=\sqrt{\frac{2}{\delta}}\varepsilon_n(t_{\text{fin}})$. 
Gathering everything together, we can express the excitation probability $p_n(t)=|\langle \text{E.S.}| \psi(t)\rangle_n|^{2}$ in terms of two dimensionless parameters
\begin{widetext}
    \begin{eqnarray}
        p_n(t)&=&1-\frac{\pi e^{-\pi \tau^2_n/8}}{2}
        \Bigg|
        \left(\frac{\sqrt{\omega^2_n(t)+\tau^2_n}+\omega_n(t)}{\sqrt{\omega^2_n(t)+\tau^2_n}}\right)^{1/2}
        \left\{\frac{M\left(\frac{i\tau^2_n}{8},\frac{1}{2},\frac{i\omega^2_n(t)}{2}\right)}{\Gamma\left(\frac{1}{2}+\frac{i\tau^2_n}{8}\right)}
        + e^{-\frac{i5\pi}{4}}\frac{\tau^2_n}{8}\frac{\omega_n(t)}{\sqrt{2}}
        \frac{M\left(\frac{1}{2}+\frac{i\tau^2_n}{8},\frac{3}{2},\frac{i\omega^2_n(t)}{2}\right)}{\Gamma\left(1+\frac{i\tau^2_n}{8}\right)}\right\} \nonumber
        \\
        &&-\frac{\tau_n}{2}\left(\frac{\sqrt{\omega^2_n(t)+\tau^2_n}-\omega_n(t)}{\sqrt{\omega^2_n(t)+\tau^2_n}}\right)^{1/2}
        \left\{\frac{e^{-\frac{i3\pi}{4}}M\left(\frac{1}{2}+\frac{i\tau^2_n}{8},\frac{1}{2},\frac{i\omega^2_n(t)}{2}\right)}{\sqrt{2} \Gamma\left(1+\frac{i\tau^2_n}{8}\right)}
        + e^{-\frac{i\pi}{2}}\omega_n(t)
        \frac{M\left(1+\frac{i\tau^2_n}{8},\frac{3}{2},\frac{i\omega^2_n(t)}{2}\right)}{\Gamma\left(1+\frac{i\tau^2_n}{8}\right)}\right\}
        \Bigg|^2,
    \end{eqnarray}
\end{widetext}
where
\begin{eqnarray*}
    \tau^2_n &=& \frac{2}{\delta}\tilde{\Delta}^2_n
    \\
    \omega^2_n(t) &=& \frac{2}{\delta}\varepsilon^2_n(t)
\end{eqnarray*}

\emph{Thermodynamic limit:}
When taking first the thermodynamic limit for the adiabatic ramps  \( t_{\text{fin}} = \frac{\lambda}{\delta}\) we have
 \( \lim_{N\rightarrow \infty}\omega^2_n(t_{\text{fin}}) = 0 \) and \( \lim_{N\rightarrow \infty}\tau^2_n(t_{\text{fin}}) = 0 \) 
    with a finite ratio $\lim_{N\rightarrow \infty}\omega^2_n(t_{\text{fin}})/\tau^2_n(t_{\text{fin}})  =\frac{ s^2_\alpha}{c^2_{\beta}}N^{2\left(\alpha-\beta\right)}$
which, at leading order in $\tau_n$ and $\omega_n$, simplies the excitation probability to
    \begin{eqnarray}
        p_n(t_{\text{fin}})&=&\frac{1}{2}\left(1 +\frac{s_\alpha}{\sqrt{s^2_\alpha+c^2_\beta N^{2(\beta-\alpha)}}}\right)
        \\
        &&-\frac{1}{2\sqrt{2\delta}}\frac{c^2_\beta N^{(2\beta-\alpha-1)}}{\sqrt{s^2_\alpha+c^2_\beta N^{2(\beta-\alpha)}}}\, .\nonumber
    \end{eqnarray}
Note that playing with the coefficients $\alpha$ and $\beta$ and/or $c_\beta, s_\alpha$ 
one can tune the instantaneous basis $\theta_n(t)$.

For the case $\beta=\alpha$ we recover the adiabatic scaling with an additional correction
    \begin{eqnarray}\label{eq:pn_correct}
        p_n(t_{\text{fin}})&=&\sin^2\left(\frac{\pi}{4}(1+\alpha)\right)-\frac{\sqrt{\pi \tau^2_n}}{4}
    \end{eqnarray}
where the coefficients $s_\alpha$ and $c_\beta$ have already been substituted by the expressions $s_\alpha=\sin (\alpha \pi / 2) \Gamma(2-\alpha)(2 \pi)^{\alpha-1}$ and $c_\alpha=\cos (\alpha \pi / 2) \Gamma(2-\alpha)(2 \pi)^{\alpha-1}$. Let us note note that the second term in Eq.\,\eqref{eq:pn_correct} scales as $\mathcal{O}(N^{\alpha-1})$ and therefore the thermodynamic limit.
When summing the contribution of all modes $n$ we can approximate it as a statistically irrelevant fraction of discrete modes plus the high-energy ones that are accumulated around $\tilde{\Delta}_n\approx 0, \varepsilon_n\approx 0$ such that:

\begin{eqnarray}
P(t_{\text{fin}})=\sum_{n}p_n(t_\text{fin})\propto N\sin^2\left(\frac{\pi}{4}(1+\alpha)\right)\,.
\end{eqnarray}


\end{document}